\documentclass[12pt]{article}
\usepackage{psfig}

\makeatletter
\def\vereq#1#2{
\lower3pt\vbox{\baselineskip1.5pt \lineskip1.5pt
\ialign{$\m@th#1\hfill##\hfil$\crcr#2\crcr\sim\crcr}}}
\makeatother

\begin{document}
\begin{titlepage}
\begin{center}
\hfill    CERN-TH/2000-379\\
\hfill FISIST/20-2000/CFIF\\
~{} \hfill hep-ph/0012289\\

\vskip 1cm

{\large \bf Split Fermions in Extra Dimensions and CP Violation}

\vskip 1cm

G. C. Branco\footnote{gbranco@thwgs.cern.ch and 
gbranco@cfif.ist.utl.pt}$^\P$
Andr\' e de Gouv\^ ea\footnote{degouvea@mail.cern.ch}
and 
M. N. Rebelo
\footnote{mrebelo@thwgs.cern.ch and
rebelo@cfif.ist.utl.pt}\footnote[5]{On leave of absence from 
Centro de F\'\i sica 
das Interac\c c\~ oes Fundamentais, CFIF, \\
Instituto Superior T\' ecnico, Av. Rovisco Pais, P-1049-001, Lisboa,
Portugal.} 
\vskip 0.05in

{\em Theory Division, CERN, CH-1211 Geneva 23, Switzerland.}
\end{center}

\vskip 3cm

\begin{abstract}

We discuss CP violation in the quark sector within a novel approach
to the Yukawa puzzle proposed by Arkani-Hamed and Schmaltz, where
Yukawa hierarchies result from localising the
Standard Model quark field wave-functions, at different positions
(in the extra dimensions) in a ``fat-brane.'' We show that 
at least two extra dimensions are necessary in order to obtain sufficient 
CP violation, while reproducing the correct quark mass spectrum and 
mixing angles.

\end{abstract}

\end{titlepage}

\newpage
\setcounter{footnote}{0}
\setcounter{equation}{0}
\section{Introduction}

Understanding the pattern of fermion masses and mixing angles is one of the 
fundamental, still unsolved, puzzles in
Particle Physics. In the Standard Model (SM), Yukawa couplings are
arbitrary free
parameters and therefore, in order to shed any light into the fermion mass
puzzle, physics beyond the SM is required. 

The traditional approach to the fermion mass issue \cite{flavour}
consists of adding spontaneously
broken flavour symmetries (gauged or global) to the SM. The symmetry and the
pattern of symmetry breaking then leads to predictions for the fermion masses and
mixing angles (or at least to correlations among them).

Recently, Arkani-Hamed and Schmaltz (AS) \cite{AS} suggested a novel approach to
the flavour puzzle, in the framework of large, extra dimensions \cite{extradims}.
They suggest that we live in a ``fat'' four-dimensional subspace (fat-brane), which
is infinite in the usual four space-time dimensions and possesses 
a finite
volume in the extra, orthogonal dimensions. In this scenario, the Higgs 
boson and
the gauge fields are free to propagate in the entire fat-brane, while fermions
have higher dimensional wave-functions which are localised in specific points 
in the extra dimensions \cite{IAKB}. Therefore, the effective four-dimensional 
Yukawa coupling
between two fermion species  turns out to be (very efficiently) suppressed
if the fermions are localised in different points, due to 
the small overlap of their respective wave-functions. Within this scenario, 
it is possible to start from a set higher-dimensional 
Yukawa couplings  of order one, 
and obtain strong and specific fermion mass hierarchies 
simply by appropriately choosing the position of the fermionic fields. It has 
been shown that this unorthodox proposal not only is capable of reproducing the 
correct pattern of quark masses and mixing angles \cite{MS}, but that it also 
has potential (and rather unique) experimental signatures \cite{AGS}.

In this paper, we address the question of CP violation in the AS scenario. 
We argue that, if the fermions are localised in different points 
in a one dimensional subspace (line),
it is not possible to obtain sufficient CP violation in order to accommodate
the current experimental data. We perform our analysis in the so-called 
nearest-neighbour-interaction (NNI) basis \cite{NNI}, and illustrate the source
of difficulties which are encountered in the context of only one 
extra dimension.\footnote{Henceforth, we refer to the number of extra dimensions 
as the dimensionality of the space where fermions are localised. This does not 
necessarily agree with the total number of large, new dimensions, which is 
required to be larger than 1 if one is to properly address the hierarchy problem
\cite{extradims}. For example, our fat-brane may live in two compact 
extra dimensions, while the fermions are localised to one-dimensional ``walls''
within the brane.} Finally, we consider models with  two extra dimensions 
and give an example where all quark masses and mixing angles are correctly 
reproduced, along with the required strength of CP violation.

This paper is organised as follows: in Sec.~2 we briefly review the AS scenario,
paying special attention to the choice of 
weak basis and its interpretation. 
In Sec.~3 we show what is required in order to obtain the correct strength of 
CP violation, while
reproducing the observed pattern of quark masses and mixing angles. 
In Sec.~4 we summarise our results and present our 
conclusions.  
  
\setcounter{footnote}{0}
\setcounter{equation}{0}
\section{The Fat-Brane Scenario}

In this section, we briefly present the fat-brane paradigm, and discuss
how it can be used to solve the fermion mass puzzle. It is instructive to start
from the $4+1$-dimensional action \cite{AS}\footnote{See \cite{Erdem} for
a more detailed treatment of the ``fat-brane'' action.}
for two fermion fields:
\begin{equation}
S\supset \int {\rm d}x^{4}{\rm d}y~\bar{Q}[i\Gamma_M\partial^M+\Phi^Q(y)]Q +
\bar{U}[i\Gamma_M\partial^M+\Phi^U(y)]U + \kappa HQ^cU,
\end{equation}
where $Q$ and $U$ are the ``(anti)quark'' fields, $H$ is the (higher-dimensional) 
Higgs field and $\kappa$ the higher-dimensional Yukawa coupling. 
$\Gamma_M$ are the $4+1$-dimensional version of the Dirac matrices 
($M=0...4$). $\Phi^{Q,U}(y)$ are potentials for the quark fields, and are 
such that the quarks are confined to specific ``points'' in the extra dimension. 
We refer to \cite{AS} for details. After expanding $Q$, $U$ and $H$ in its 
``normal modes'' (properly normalised), the Yukawa part of the action for the 
zero-modes is
\begin{equation}
S_{\rm Yuk}=\int {\rm d}x^{4}~\kappa h(x)q(x)u(x)\int {\rm d}y~\phi_q(y)\phi_u(y),
\end{equation}
where $\phi_{q,u}(y)$ are the fifth dimensional wave-functions for the $Q$ and
$U$, and $h(x)$, $q(x)$, $u(x)$ are the four-dimensional Higgs, $q$ and $u$
fields, respectively. It is assumed that the Higgs zero-mode is independent
of $y$, the extra dimension. Assuming $\phi_{q,u}(y)$ to be Gaussians centred at
$l_q$ and $l_u$, respectively, with width $1/(\sqrt{2}\mu)$ \cite{AS}, 
\begin{equation}
\int {\rm d}y~\phi_q(y)\phi_u(y)=
\frac{\sqrt{2}\mu}{\sqrt{\pi}}\int {\rm d}y~e^{-\mu^2(y-l_q)^2}e^{-\mu^2(y-l_u)^2}
=e^{-\mu^2 (l_q-l_u)^2/2},
\label{exp_factor}
\end{equation} 
where the second equality is valid when the thickness of the brane (range of $y$)
is significantly larger than $1/\mu$ and $l_{q,u}$.
The effective four-dimensional Yukawa coupling is then of the form 
$\lambda=\kappa~e^{-\mu^2 (l_q-l_u)^2/2}$. The exponential factor obtained
from Eq.~(\ref{exp_factor}) is of key importance, since it allows for 
(exponentially) suppressed Yukawa couplings even when the original 
higher-dimensional couplings are of order one, if the different quark fields
are confined at different points in the extra dimension. Such mechanism is 
capable of not only generating very small Yukawa couplings, but may also be used
to suppress dangerous higher-dimensional operators  mediating 
proton decay, $K^0\leftrightarrow \bar{K}^0$ mixing, etc \cite{AS},
which normally plague theories with a small fundamental scale.  
 
A pertinent question is whether there is any geometrical configuration 
of quark fields which fits all quark masses and mixing angles. This issue
was addressed in \cite{MS}, and the answer is positive. Therefore, without
assuming any flavour symmetry ({\it i.e.}\/ all $\kappa$ of order one), it
is possible to accommodate the observed pattern of
fermion masses and mixing angles simply by 
appropriately placing each quark field in a different position.   

We address the issue of the quark Yukawa matrix in more detail in order
to discuss a few relevant points. First, we choose a basis where the weak
charged current is diagonal, such that 
\begin{equation}
{\cal L}_{\rm Yuk}=\lambda_u^{ij}Q_iU_jH + \lambda_d^{ij}Q_iD_jH^{*},
\end{equation}
where $Q_i$ are the quark doublets, and $U_i$, $D_i$ the up-type and down-type
antiquark singlets, respectively ($i,j=1,2,3$ for the three families). According
to Eq.~(\ref{exp_factor}), 
$\lambda_{u}^{ij}=\kappa^{ij}\exp(-\mu^2(l_{q_i}-l_{u_j})^2/2)$ (repeated indexes
not summed over). The same holds for $\lambda_{d}^{ij}$ with $l_{u_j}$ replaced by
$l_{d_j}$. 

Therefore, as was mentioned previously, if all $\kappa^{ij}$ are of order one,
the whole texture of the Yukawa matrix is dictated by the relative distance
of $q$'s and $u$'s, $d$'s. An important comment is that the exponential factors
dramatically affect the otherwise arbitrary moduli of the 
higher-dimensional Yukawa couplings, but
they do nothing to potentially large complex phases. Therefore, in general terms,
each element of $\lambda_{(u,d)}^{ij}$ is accompanied by arbitrary, unsuppressed
complex phases.

Another interesting point is that, since in the AS framework quark 
fields are localised in different places, families are distinguishable, at
least in principle, even in the limit where all Yukawa couplings vanish.
As a result, there is no freedom to rotate the fermion fields
in family space (no $U(3)^5$ global symmetry). Ultimately a weak basis (WB)
will be dictated by the localising mechanism for the fermion fields.
However, one can still refer to different choices of WB which should be
understood as corresponding to different assumptions about the 
underlying physics.

\setcounter{equation}{0}
\setcounter{footnote}{0}
\section{Realistic Quark Masses and CP Violation}

Within the AS approach to the Yukawa puzzle the choice of WB 
plays a crucial r\^ ole. It
is worthwhile to comment on what WB
choices are more appropriate for  the fat-brane scenario. 

Let us recall that in the standard approach to the fermion mass problem,
flavour symmetries added to the SM lead to: $i$-zeros in the Yukawa matrices,
resulting from terms forbidden by the flavour symmetry and/or
$ii$-equality and/or relations
between elements in the Yukawa matrices, which are connected by the flavour
symmetry.
In the AS approach, as discussed in the previous section, it is quite natural to
obtain effective zeros in the Yukawa matrices, since they correspond to elements
which 
connect fermions which are very ``far'' from one another. On the other hand, 
equalities or specific relations among elements of the Yukawa matrices are not
``natural'' in the AS scheme -- they require fine-tuned choices for the positions
of the fermion fields in the extra dimensions. 

From the above considerations, one is led to conclude that, within the
fat-brane
framework, the most convenient WB are those which contain a large number 
of zeros, while bases where the fermion mass matrices are either symmetric or
Hermitian are not adequate. Two especially interesting bases are the 
nearest-neighbour-interaction (NNI) basis,
\begin{equation}    
M_d=\left(\matrix{0 & a & 0 \cr a' & 0 & b \cr 0 & b' & c} \right); \hspace{1cm} 
M_u=\left(\matrix{0 & d & 0 \cr d' & 0 & e \cr 0 & e' & f} \right), 
\label{NNI_basis}
\end{equation}
and the triangular (T) basis,
\begin{equation}    
M_d=\left(\matrix{g & k & l \cr 0 & m & n \cr 0 & 0 & p} \right); \hspace{1cm}
M_u=\left(\matrix{m_u & 0 & 0 \cr 0 & m_c & 0 \cr 0 & 0 & m_t} \right), 
\end{equation}
where $M_d$ and $M_u$ are the down-type and up-type mass-matrices (after
electroweak symmetry breaking) with complex entries.

It should be emphasised that, in the SM, both the NNI \cite{NNI}
and T forms\footnote{The NNI basis should not be
confused with the Fritzsch ansatz \cite{Frit} which consists of taking
the NNI form together with Hermiticity. NNI with this additional requirement 
does lead to physical predictions which have already been ruled out 
by experiment due to the observed large top quark mass.}
correspond to a choice of WB, in the sense that starting from arbitrary 
mass matrices $M_u$ and $M_d$ one can always perform transformations of the type
\begin{eqnarray}
M_u\rightarrow M_u'=W_L^{\dagger}M_uW_R^u, \\ 
M_d\rightarrow M_u'=W_L^{\dagger}M_dW_R^d,
\end{eqnarray}
where the $W$'s are unitary matrices, which transform $M_u$ and $M_d$ into 
the NNI or the T form
while leaving the weak currents diagonal in flavour space
(note that $W_L$ is the
same matrix in $M_u$ and $M_d$).

\subsection{One Extra Dimension}

Recently, Mirabelli and Schmaltz \cite{MS} (MS) performed 
what they referred to as a ``brute force'' scan
over the parameter space in the framework of one extra dimension, and found the 
following mass matrices \cite{MS}:
\begin{eqnarray}    
M_d=\left(\matrix{0 & 16.974 & 0 \cr 14.510 & 0 & 123.42 \cr 0 & 1373.2 & 2370.2} 
\right)~{\rm MeV} & \nonumber \\ 
M_u=\left(\matrix{1.7630 & 0 & 0 \cr 0 & 576.06 & 2.7882\times 10^{-3}
\cr 5902.8 & 0 & 165900} 
\right)~{\rm MeV} &\simeq \left(\matrix{m_u & 0 & 0 \cr 0 & m_c & 0 \cr 0 & 0 &
m_t} 
\right),
\label{MS_matrix}
\end{eqnarray}
where the zeros correspond to elements strongly suppressed by exponential factors.
These matrices lead to the correct values for quark masses and mixing angles. 
We now address the issue of CP violation, which was not discussed in \cite{MS}. 
In order to do so, we assume that all entries in $M_u$ and $M_d$ are complex, 
with arbitrary
phases (as discussed in Sec.~2). From Eq.~(\ref{MS_matrix}), one obtains,
to a very good approximation,
\begin{eqnarray}
H_u\equiv M_uM_u^{\dagger} \simeq
\left(\matrix{m_u^2 & 0 & 0 \cr 0 & m_c^2 & 0 \cr 
0 & 0 & m_t^2} \right), \nonumber \\
H_d\equiv M_dM_d^{\dagger}=\left(
\matrix{H^{d}_{11} & 0 & H^{d}_{13} \cr 0 & H^{d}_{22} & H^{d}_{23} 
\cr H^{d*}_{13} & H^{d*}_{23} & H^{d}_{33}} \right).
\label{MS_mass-squared}
\end{eqnarray} 

For any arbitrary set of quark mass matrices, in the weak basis where $H_u$ is 
diagonal, all $|H_d|_{ij}$ have physical meaning and can be expressed as a function
of quark masses and mixing angles. In particular one has
\begin{equation}
H^d_{12}=m_d^2 V_{ud}V_{cd}^*+m_s^2V_{us}V_{cs}^*+m_b^2V_{ub}V_{cb}^*,
\label{H12}
\end{equation}
where $V_{ij}$ are elements of the Cabibbo-Kobayashi-Maskawa (CKM) matrix.
Since in the MS ansatz $H^d_{12}$ vanishes, from  
Eq.~(\ref{H12}) together with unitarity of the CKM matrix, one obtains
\begin{equation}
\frac{m_s^2-m_d^2}{m_b^2-m_d^2}=\frac{|V_{ub}||V_{cb}|}{|V_{us}||V_{cs}|}
\label{ms}
\end{equation}  
and
\begin{equation}
{\rm arg}(V_{ub}V_{cs}V_{us}^*V_{cb}^*)=\pi.
\label{pi}
\end{equation}
Eq.~(\ref{ms}) corresponds to  relation (7.12) of reference \cite{MS}, 
which was derived in a slightly 
different way, while Eq.~(\ref{pi}) implies that there is no CP violation in
the solution found in \cite{MS} (namely, Eq.~(\ref{MS_matrix})). An alternative 
way of showing that CP is not violated is by evaluating the following WB
invariant
\begin{equation}
Tr[H_u,H_d]^3=6i(m_u^2-m_c^2)(m_t^2-m_c^2)(m_t^2-m_u^2)
\mbox{Im} (H^d_{12}H^d_{23}H^{d*}_{13})=0,
\end{equation}
where the first equality is obtained in the WB where $H_u$ is diagonal while
the second equality results from the fact that $H^d_{12}$ vanishes in this 
ansatz. The vanishing of the above invariant is a necessary and sufficient
condition  \cite{CPV_cond} for $CP$-invariance
in the SM. Note that $M_d$ in
Eq.~(\ref{MS_matrix}) has the NNI form, and this is why $H^d_{12}$ vanishes.
This, in
turn, implies that Eq.~(\ref{ms}) has a much larger range of applicability
than the ansatz presented in \cite{MS}, being valid for any model where $M_d$ 
has the NNI form while $H_u$ is diagonal.

One may wonder whether our conclusion about the absence
of CP violation in the MS ansatz is affected by the
fact that $M_u$ in Eq.~(\ref{MS_matrix}) is not exactly diagonal. It can be
readily seen that taking $M_u$ diagonal is a very good approximation (as argued
in \cite{MS}), so that when one considers the effect of the off-diagonal
terms, the resulting strength of CP violation is always  
too small to account for the
experimentally observed value of $\epsilon_K$ \cite{PDG}. Indeed, if
one allows for complex arbitrary phases in the entries of $M_d$, $M_u$
in Eq.~(\ref{MS_matrix}), one can readily derive an upper bound on the 
strength of CP violation, namely 
$J\equiv |\mbox{Im} (V_{ub}V_{cs}V_{us}^*V_{cb}^*)|\leq 5 \times 10^{-9}$.
This is to be compared to  
$J\simeq 10^{-5}$, required by the experimental value of $\epsilon_K$ . 

We now proceed with a systematic study of fermion mass matrices in the AS scheme.
We shall work in the NNI basis defined at the beginning
of this section by Eq.~(\ref{NNI_basis}) and assume that 
the entries of $M_u$ and
$M_d$ have arbitrary phases. 

At this point one should mention that although the MS example of
Eq.~(\ref{MS_matrix}) is not explicitly written in the NNI basis, a simple 
WB transformation consisting of the permutation of the first two columns
transforms their ansatz into the NNI form. Geometrically this transformation 
corresponds to the interchange of positions between $U_1$ and $U_2$ in
the extra dimension. The relative positions of the quark wave-functions 
giving rise to the NNI basis in one dimension are depicted in 
Fig.~1, which is to be compared to Fig.~5 in \cite{MS}.

\begin{figure}
\label{fig:1-dim}
    \centerline{
    \psfig{file=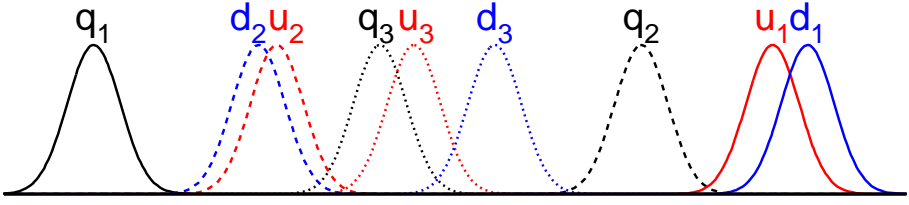,width=1\columnwidth}
}
    \caption{Quark wave-functions in the one extra dimension solution presented by
 MS, rotated to the NNI-form.}
\end{figure}

Since, in the SM, there are no right-handed currents,
the mass spectrum and the CKM-matrix only depend on $H_u\equiv M_uM_u^{\dagger}$
and $H_d\equiv M_dM_d^{\dagger}$, which have the following
form in the NNI basis:
\begin{equation}
H_{(u,d)} = \left(\matrix{H^{(u,d)}_{11} & 0 & H^{(u,d)}_{13} 
\cr 0 & H^{(u,d)}_{22} & H^{(u,d)}_{23} 
\cr H^{(u,d)*}_{13} & H^{(u,d)*}_{23} & H^{(u,d)}_{33}} \right).
\end{equation}
By making a transformation of the type 
\begin{equation}
H_u\rightarrow K^{\dagger}H_uK; \hspace{1cm} H_d\rightarrow K^{\dagger}H_dK,
\end{equation}
where $K$ is a diagonal unitary matrix, it is possible to eliminate all complex
phases from $H_u$, while the off diagonal elements of
$H_d$ still have arbitrary phases. In this case, both matrices
are diagonalized in the following way,
\begin{eqnarray}
O_u^{\dagger}~H_u~O_u={\rm diag}(m_u^2,m_c^2,m_t^2), \label{Du}\\
O_d^{\dagger}K'^{\dagger}~H_d~K'O_d={\rm diag}(m_d^2,m_s^2,m_b^2),
\label{Dd}
\end{eqnarray} 
where the $O$'s are orthogonal matrices and $K'$ is a unitary diagonal matrix
whose r\^ ole is to eliminate the phases in $H_d$.
Without
loss of generality, one may choose $K'={\rm diag}(1,e^{i\phi},e^{i\sigma})$.

The CKM matrix is, therefore,  given by
\begin{equation}
V_{CKM}=O_u^{\dagger}K'O_d.
\label{VCKM}
\end{equation}
The relevant elements for our discussion can be explicitly written as
\begin{eqnarray}
V_{us}=O^u_{11}O^d_{12}+O^u_{21}O^d_{22}e^{i\phi}
+O^u_{31}O^d_{32}e^{i\sigma}, \label{Vus} \label{forfig} \\
V_{ub}=O^u_{11}O^d_{13}+O^u_{21}O^d_{23}e^{i\phi}
+O^u_{31}O^d_{33}e^{i\sigma}, \\
V_{cb}=O^u_{12}O^d_{13}+O^u_{22}O^d_{23}e^{i\phi}
+O^u_{32}O^d_{33}e^{i\sigma}, \\
V_{cs}=O^u_{12}O^d_{12}+O^u_{22}O^d_{22}e^{i\phi}
+O^u_{32}O^d_{32}e^{i\sigma}, \\
V_{td}=O^u_{13}O^d_{11}+O^u_{23}O^d_{21}e^{i\phi}
+O^u_{33}O^d_{31}e^{i\sigma}. \label{Vtd}
\end{eqnarray} 

   In order to check whether it is possible to obtain, with only one extra
dimension, sufficient CP violation, we have done a series of guided 
trials using Eqs.~(\ref{Vus})-(\ref{Vtd})
and taking into account the observed quark mass spectrum.
Soon it became apparent that this it is
not
possible with only one extra dimension. This can be understood by the
following argument. Since we are working in the NNI basis (where 
$H^d_{12}$ vanishes), the only way to obtain sufficient CP violation
is by having, in the same basis, $H_u$  significantly deviating from
the diagonal. This in turn requires more proximity between the $U_i$'s
and $Q_j$'s. Once the $U_i$'s get close enough to the $Q_j$'s there
is no room to place the $D_k$'s along the same line (since the $Q_j$'s
are now closer to each other) while at the same time obtaining the
correct masses and mixing. Indeed with only one extra dimension,
we have not found any solution significantly different from the one
proposed by MS, which would correctly reproduce the quark masses
and mixing angles. This confirms the uniqueness of the MS solution
(modulo WB transformations on right handed quark fields), in 
the case of only one extra dimension.

\subsection{Two Extra Dimensions}
In the search for a solution with two extra dimensions,
we shall continue to work in the NNI basis and look for configurations 
of the quark fields where the CKM matrix arising from Eq.~(\ref{VCKM}),
although dominated by $O_d$, receives a significant contribution
from $O_u$. As we have previously argued, for NNI quark mass matrices,
the existence of a non-negligible contribution from $O_u$ is essential
in order to be able to generate sufficient CP violation. It is clear
from Eqs.~(\ref{Du}),(\ref{Dd}) that the orthogonal matrices $O_u$, 
$O_d$ are determined by the relative positions of the quark
fields in the extra dimension space, while the phases $\phi$ and 
$\sigma$ are free parameters. Both the modulus and the argument of
$V_{us}$ crucially depend on the phase $\phi$, while $\sigma$ does 
not play much of a r\^ ole in the determination of $V_{us}$, due to the
smallness of the factor $O^u_{31}O^d_{32}$ (as will be seen later). 
On the other hand, for the elements $V_{ub}$ and $V_{cb}$, 
both $\phi$ and $\sigma$ potentially play an important r\^ ole. 

\begin{figure}
\label{fig:2-dim}
    \centerline{
    \psfig{file=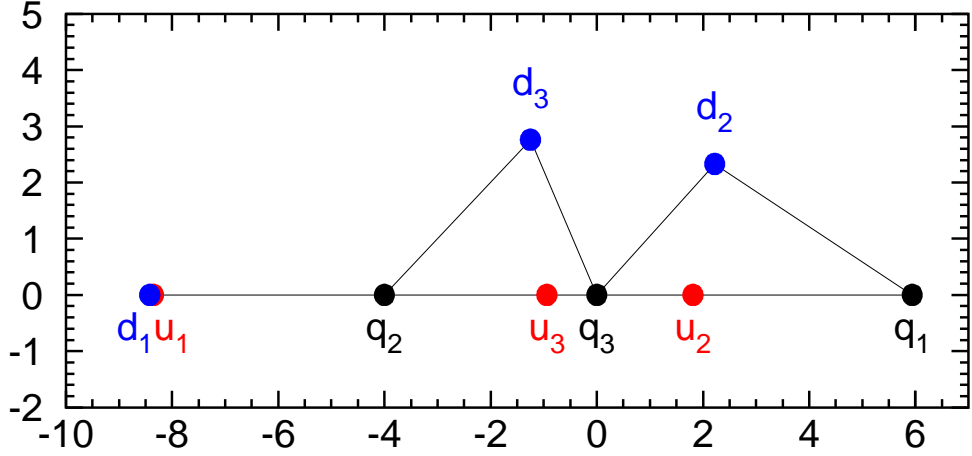,width=1\columnwidth}
}
    \caption{Locations of the quark wave-functions corresponding to
Eq.~(\ref{2d_points}). 
Distances are measured in units of $\mu^{-1}$ (see text).
The lines indicate distances which are dictated by the nonzero entries
of $M_d$, $M_u$ given in Eq.~(\ref{choice}). Note that once the 
$u_i$'s and $q_i$'s are placed on the same straight line, $d_2$ and $d_3$
are forced into the second dimension.}
\end{figure}

We have found an interesting set of locations
for the quark fields which leads to the right spectrum of quark masses
and pattern of mixing angles, while allowing for the right strength of 
CP violation. The locations of the quark fields in the two extra 
dimensions are depicted in Fig.~2. Explicitly,
\begin{equation}
q_i = \frac{1}{\mu} \left( \begin{array}{c}5.941; 0 \\  
-4.008; 0 \\ 
0;0  \end{array} \right), 
u_i = \frac{1}{\mu}  \left( \begin{array}{c}-8.347; 0 \\
1.815; 0 \\
-0.941; 0 \end{array} \right), 
d_i =  \frac{1}{\mu} \left( \begin{array}{c}-8.421; 0 \\
2.219; 2.332 \\
-1.253; 2.767 \end{array} \right),
\label{2d_points}
\end{equation}
which lead to the following masses matrices, assuming $\kappa^{ij} v=1.5m_t$, 
for all $i$ and $j$ (see \cite{MS} for details regarding this choice):
\begin{eqnarray}
M_d=\left( \begin{array}{ccc} 0 & 16.112 & 0 \\
14.690 & 0 & 121.77 \\ 0 & 1400 & 2467.8 \end {array} \right) 
\; \mbox{MeV}, \nonumber \\
M_u=\left( \begin{array}{ccc} 0 & 50.0 & 0 \\
20.3 & 0 & 2258 \\ 0 & 48 000 & 160 000 \end {array} \right)
\; \mbox{MeV},
\label{choice}
\end{eqnarray}
where the zeros correspond to strongly suppressed matrix elements.
From Eq.~(\ref{choice}) one readily obtains $O_d$ and $O_u$. The allowed
range of values for $\phi$ is essentially dictated by the 
experimental value of $| V_{us} |$,
\begin{equation}
| V_{us} | = 0.219 \; \mbox{to} \; 0.226.
\label{range}
\end{equation}
Indeed from Eq.~(\ref{choice}) it follows that
\begin{eqnarray}
O^u_{11} = 0.9973; \; O^u_{21} = 0.0735; \; O^u_{31} = -0.0010; 
\nonumber \\
O^d_{12} = 0.2157; \; O^d_{22} = -0.9758; \; O^d_{32} = 0.0358.
\label{OuOd}
\end{eqnarray}
From Eqs.~(\ref{Vus}), (\ref{range}), and (\ref{OuOd}), it follows 
that $\phi$ is constrained to be in the range
\begin{equation}
         83.6^{\circ} \leq \phi \leq 89.4^{\circ},
\end{equation}
while $\sigma$ does not play much of a r\^ ole in the determination 
of $| V_{us} | $, since $| O^u_{31}O^d_{32}| \sim 10^{-5}$.
This situation is depicted in Fig.~3.
We choose $\phi$ in the first quadrant in order to obtain the
appropriate sign for $\mbox{Im} (V_{ub}V_{cs}V_{us}^*V_{cb}^*)$.
\begin{figure}
\label{fig:Vus}
    \centerline{
    \psfig{file=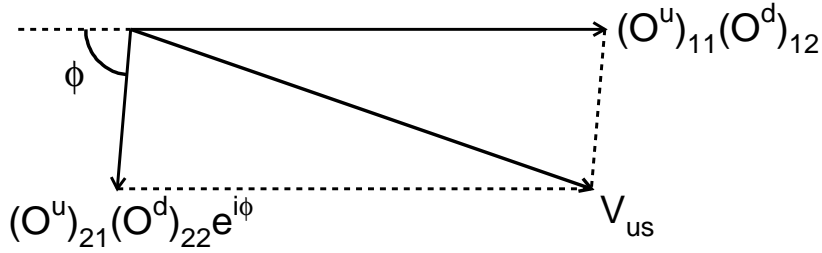,width=1\columnwidth}
}
    \caption{$V_{us}$ in the complex plane (see Eq.~(\ref{forfig}) and 
text for details).}
\end{figure}
 
It is remarkable that in order to obtain the correct value of 
$| V_{us} | $ one is led to a large value of $\phi$ which
in turn is crucial to have a sufficient amount of CP violation.
For $\phi = 85^{\circ}$, $\sigma = 0^{\circ}$, we have obtained
\begin{equation}
| V_{CKM} | = \left( \begin{array}{ccc} 0.9753  & 0.2208 & 0.0034 \\
0.2205 & 0.9746 & 0.0384 \\ 0.0108 & 0.0370 & 0.9993 \end{array} \right)
\end{equation} 
The strength of CP violation can be readily evaluated:
\begin{equation}
J\equiv |\mbox{Im} (V_{ub}V_{cs}V_{us}^*V_{cb}^*)|\simeq 2.2 \times 10^{-5}.
\end{equation}

The values of the quark masses implied by Eq.~(\ref{choice}) are
\begin{eqnarray}
& m_u=1.5 \; \mbox{MeV}, m_d=3.2 \; \mbox{MeV}, m_s=63.3 \; \mbox{MeV},
\nonumber \\
& m_c=651 \; \mbox{MeV}, m_b=2839 \; \mbox{MeV}, m_t=167 059 \; \mbox{MeV},
\end{eqnarray}
corresponding to the quark masses evaluated at the common scale $m_t$. 
These are in agreement with the experimental values of the quark masses 
(computed in the $\overline {\mbox{MS}}$
renormalization scheme, with up, down, and strange quark masses 
evaluated at a scale of 2 GeV and the others evaluated at a scale equal to their 
$\overline {\mbox{MS}}$ mass), as compiled by the Particle Data Group \cite{PDG}:
\begin{eqnarray}
& m_u=1.5 \; \mbox{to} \; 5 \; \mbox{MeV}, m_d=3 \; \mbox{to} \; 9 \; \mbox{MeV}, 
m_s=75  \; \mbox{to} \; 170 \; \mbox{MeV}, \nonumber \\ 
& m_c=1150 \; \mbox{to} \; 1350 \; \mbox{MeV}, 
m_b=4000 \; \mbox{to} \; 4400 \; \mbox{MeV},\nonumber \\
& m_t=166 000 \pm 5 000 \; \mbox{MeV},
\end{eqnarray}
when one takes the effect of renormalization group running into account. 
In order to do this  
we use scaling factors $\eta_i$ such that  
$\eta _i \equiv m_i(m_i)/m_i(m_t)$ for $i=c,b,t$  and
$\eta _i \equiv m_i(\mbox{2 GeV)}/m_i(m_t)$  for $i=u,d,s$.
These have been computed to three loops in QCD and one loop in QED \cite{etas},
and are given by
\begin{equation}
\eta _u=1.84, \; \eta _d=1.84, \; \eta _s=1.84, \; \eta _c=2.17, 
\; \eta_b=1.55,\; \eta _t=1.00.
\end{equation}
One comment which should be made at this point is that the fact that $H_u$ in the 
example discussed is not diagonal leads to a correction of order 10\% in the 
value of $m_s$ implied by Eq.~(\ref{ms}).     

Finally, we would like to comment on the possibility of having sufficient
CP violation with the triangular basis. As far as obtaining the correct
mass spectrum and mixing angles, it was already pointed out \cite{MS} that
the T basis requires at least two extra dimensions. Although we have not
studied this case in detail, we do not anticipate, a priori, any difficulty
in obtaining the right amount of CP violation as well.

\setcounter{equation}{0}
\setcounter{footnote}{0}
\section{Summary and Conclusions}       
A novel approach to the fermion mass puzzle was recently proposed 
in the context of large extra dimensions by Arkani-Hamed and Schmaltz
(AS) \cite{AS} and further analysed by Mirabelli and Schmaltz
\cite{MS}. In this approach, the fermion mass hierarchy and 
mixing pattern are a consequence of the fact that different 
fermionic fields are localised in slightly different points in the higher
dimensional space. Although at this stage there is no fundamental
understanding of why various fermions would be localised at different
positions, the proposed paradigm provides a new and drastically different 
approach to the Yukawa puzzle.

In this paper we addressed the issue of CP 
violation in the AS framework. 
First, we  considered a specific example proposed 
by Mirabelli and Schmaltz (MS)
which leads to the correct quark mass spectrum and mixing angles, in
the framework of one extra dimension. We  showed that in the MS example,
even if one allows for complex mass matrices with arbitrary phases, one
can never generate sufficient CP violation through the Kobayashi-Maskawa 
mechanism. 
We performed our analysis in the nearest-neighbour-interaction (NNI) 
weak basis (WB), which is well suited for understanding fermion masses 
in the AS scheme. In fact, we showed that  although the MS quark 
mass matrices
are not written in the NNI basis, they can be transformed into that form 
by a simple WB transformation interchanging two columns of the up quark
mass matrix. 
We  also showed that a specific relation between quark masses and
mixing angles (see Eq.~(\ref{ms})) obtained by MS, can be derived in a more
general framework which goes beyond the specific MS ansatz.
A systematic search was conducted with one extra dimension 
and quark mass matrices
written in the NNI weak basis and it was shown that again it is not possible to
generate sufficient CP violation. We then considered the case of two extra
dimensions and constructed an example where the location of the 
fermion fields leads to the correct mass spectrum and mixing angles,
while allowing for the generation of sufficient CP violation to account for
the experimental value of $\epsilon_K$. In the example  considered,
one is led to a striking connection between the value of
$| V_{us} |$ and the strength of CP violation. 
 
Finally, we would like to comment on lepton masses and mixing. 
Although the fat-brane approach can also provide an 
understanding of the hierarchy  of lepton masses, 
the observed large mixing in the neutrino sector, together with the fact that
neutrinos can also be Majorana particles represents a further challenge
\cite{new} to the AS approach.

\section*{Acknowledgements}

We thank Nima Arkani-Hamed and Martin Schmaltz for comments and words
of encouragement. GCB and MNR thank the CERN Theory Division for  
hospitality.

\end{document}